# OPEN CORES FOR DIGITAL SIGNAL PROCESSING

# NÚCLEOS LIBRES PARA PROCESAMIENTO DIGITAL DE SEÑALES

# OPEN CORES FOR DSP

Juan Camilo Valderrama-Cuervo: Ingeniero Electrónico de la Universidad del Quindío. Secretaría de Transporte y Tránsito del Municipio de Envigado.

Alexander López-Parrado[1]: Ingeniero Electrónico de la Universidad del Quindío, Magíster en Ingeniería Electrónica de la Universidad del Valle, Candidato a Doctor en Ingeniería Eléctrica y Electrónica de la Universidad del Valle. Profesor Asistente del Programa de Ingeniería Electrónica e Investigador del Grupo de Investigación GDSPROC de la Universidad del Quindío.

## ABSTRACT

This paper presents the design and implementation of three System on Chip (SoC) cores, which implement the Digital Signal Processing (DSP) functions: Finite Impulse Response (FIR) filter, Infinite Impulse Response (IIR) filter and Fast Fourier Transform (FFT). The FIR filter core is based on the symmetrical realization form, the IIR filter core is based on the Second Order Sections (SOS) architecture and the FFT core is based on the Radix $2^2$ Single Delay Feedback ($R2^2SDF$) architecture. The three cores are compatible with the Wishbone SoC bus and they were described using generic and structural VHDL. In system hardware verification was performed by using an OpenRisc-based SoC synthesized on an Altera FPGA, the tests showed that the designed DSP cores are suitable for building SoC based on the OpenRisc processor and the Wishbone bus.

**Keywords**: Digital signal processing, digital filters, finite impulse response filters,

---

[1] Autor de correspondencia: Carrera 15 Calle 12 Norte, Universidad del Quindío, Código Postal 630004, Bloque de Ingeniería, Tercer Piso, CEIFI, Armenia, Colombia. Tel: 3108282506, correo electrónico: parrado@uniquindio.edu.co, alexander.lopez.parrado@correunivalle.edu.co.

infinite impulse response filters, fast Fourier transforms, system-on-chip, open source hardware, open RISC processor, Wishbone bus.

**RESUMEN**

Este artículo presenta el diseño e implementación de tres núcleos para sistemas en un solo chip (SoC) que implementan las funciones de procesamiento digital de señales (DSP): filtro de respuesta finita al impulso (FIR), filtro de respuesta infinita al impulso (IIR) y transformada rápida de Fourier (FFT). El núcleo de filtro FIR está basado en la estructura simétrica, el núcleo de filtro IIR está basado en la arquitectura de secciones de segundo orden (SOS) y el núcleo de la FFT está basado en la arquitectura base $2^2$ *Single Delay Feedback* (R2$^2$SDF). Los tres núcleos son compatibles con el bus para SoC *Wishbone* y fueron descritos usando VHDL estructural y genérico. Se realizó una verificación en hardware usando un SoC basado en el procesador OpenRISC y sintetizado en un FPGA de Altera, las pruebas mostraron que los núcleos DSP son apropiados para construir un SoC basado en el procesador OpenRISC y el bus *Wishbone.*

**Palabras clave**: Procesamiento digital de señales, filtros digitales, filtros de respuesta al impulso finita, filtros de respuesta al impulso infinita, transformada rápida de Fourier, sistemas en un solo chip, hardware de código abierto, procesador OpenRISC, bus *Wishbone*.

**INTRODUCTION**

Today's technology uses heavily Digital Signal Processing (DSP) on its applications and from the past 20 years [1] these applications have been growing up because the performed improvements to digital integrated circuits in speed, integration capabilities and power consumption. The increased speed of integrated circuits allows real time processing of signals with higher bandwidths such as the ones used in communication systems [1].

Nowadays there are Digital Signal Processors (DSPs) devices specifically designed for DSP which perform real time filtering, Fourier transforms, Wavelet transforms, or encoding processes on audio and video signals. Nevertheless, the parallel nature of DSP algorithms has motivated the research interest to hardware solutions based on reconfigurable targets such as the Field Programmable Gate Arrays (FPGAs); these solutions have demonstrated improvements in speed and power consumption compared with the DSPs-based ones [2].

There are several FPGA-based DSP solutions which are developed by private corporations such as Altera and Xilinx. These solutions include FIR filtering cores [3][4], FFT cores [5][6] among others; however these cores have expensive licenses for commercial use or they can be used for free only for academic purposes.

Nonetheless, a new open source hardware development model inspired from open source software models has been deployed from the last ten years. This model has been supported by communities like OpenCores which develops open source hardware under the Lesser General Public License (LGPL). OpenCores community has remarkable products as the OpenRISC processor core [7] and the Wishbone bus specification [8] which jointly allow the development of SoC hardware. However, OpenCores community lacks of fully parameterizable DSP cores compatible with the Wishbone bus.

By considering previous ideas we developed the cores FIR filter, IIR filter and FFT under the LGPL license which are compatible with the Wishbone bus and they allow the development of DSP-SoC based on the OpenRISC processor [9]. The FIR filter core is based on the symmetrical architecture [1][2], the IIR core is based on the SOS architecture [1][2] and the FFT core is based on the $R2^2$SDF architecture [10]. The three cores were described using generic and structural VHDL and targeted to an Altera FPGA device.

This paper is organized as follows: First, some theoretical concepts about DSP and Wishbone bus are described, then the design of the DSP cores architecture is presented and its functional blocks are described, later the in-system hardware verification results are discussed, and finally the conclusion and the acknowledgements are presented.

# THEORETICAL BACKGROUND

In this section we present some theoretical concepts about the DSP functions we implemented and the SoC bus Wishbone.

## FIR Filters

FIR filters are discrete Linear Time Invariant (LTI) systems which have a finite duration impulse response $h[n]$, when $h[n]$ is symmetrical the FIR filter has linear phase [1] leading to a constant group delay. Practical implementations of FIR filters are always stable because of their non-recursive nature. Eq. **(1)** shows the direct realization of a FIR filter.

$$y[n] = \sum_{k=0}^{N-1} h[k]x[n-k] \qquad (1)$$

Here, $x[n]$ is the input signal, $y[n]$ is the output signal and $N$ is the length of the impulse response $h[n]$. There are several realization forms for FIR filters [1]; the direct form, the symmetrical form, and the transpose form are the most used [1]. In the case of hardware implementations the transpose form has the shortest critical path and it is less sensitive to the round-off errors when fixed point arithmetic is used [1][2]. Figure 1 shows the transpose realization form for a FIR filter with impulse response of length $N$. From Figure 1 it can be seen that the critical path of the transpose realization form is determined by the combinatorial elements multiplier and adder. The transpose realization form uses $N-1$ registers.

## IIR Filters

IIR filters are discrete Linear Time Invariant (LTI) systems which have an infinite duration impulse response. Practical implementations of IIR filters can become unstable because of their recursive nature [1]. Eq. (2) shows the direct realization of a $(M-1)$-th order IIR filter.

$$y[n] = \sum_{k=0}^{M-1} b_k x[n-k] - \sum_{k=1}^{M-1} a_k x[n-k] \qquad (2)$$

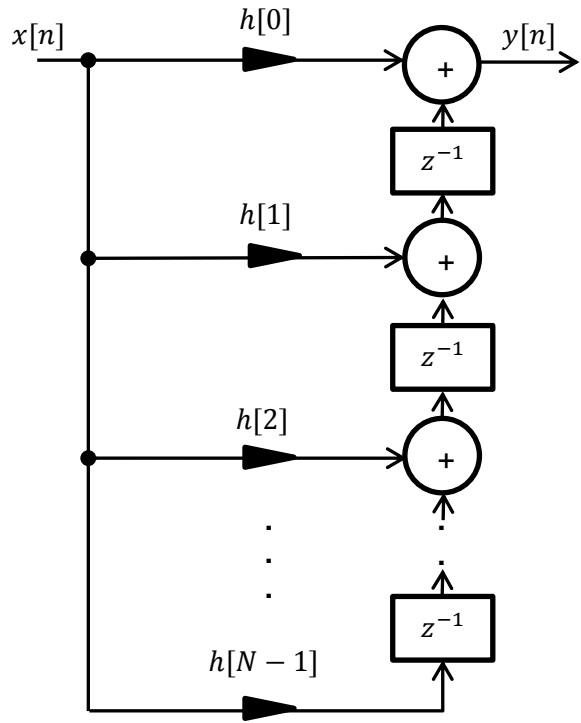

Figure 1: Transpose realization form for a FIR filter of length $N$.

Here, $b_k$ is the coefficients set for the non-recursive part, $a_k$ is the coefficients set for the recursive part, $x[n]$ is the input signal, and $y[n]$ is the output signal. There are several realization forms for IIR filters [1]; the direct form, the type II form, the transpose type II form, and the SOS form are the most used [1]. The transpose type II form has the shortest critical path; nonetheless the SOS form is less sensitive to the round-off errors when fixed point arithmetic is used [1][2]. In the case of hardware implementations the SOS form has good stability for high order filters, and the critical path is minimized by using the transpose type II form for each second order section. Figure 2 shows the transpose type II realization form for a single second order section. Here, $N_{sect}$ is the number of second order sections and $G$ is the total gain after the SOS decomposition [1], thus each second order section has a gain of $\sqrt[N_{sect}]{G}$. The whole IIR filter is composed by a cascade of $N_{sect}$ second order sections as the shown in Figure 2. From Figure 2 it can be seen that the critical path of the transpose type II realization form is determined by two multipliers and two adders.

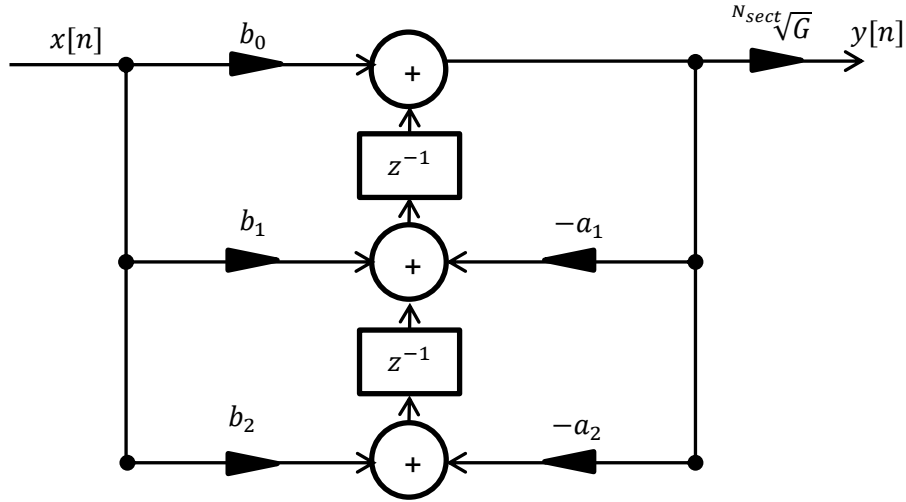

Figure 2: Transpose type II realization form for second order section.

*Fast Fourier Transform*

The FFT is an algorithm which efficiently computes the Discrete Fourier Transform (DFT) of a discrete time signal [1][2]. The DFT of a signal $x[n]$ is shown in Eq. (3)

$$X[k] = \sum_{n=0}^{N-1} x[n] e^{-\frac{i2\pi k n}{N}} \tag{3}$$

According to the used radix, FFT algorithms can be radix-2, radix-4, radix-$2^2$, radix-8, mixed-radix, split-radix [2][5][6][10], among others. The radix-$2^2$ algorithms have become popular for hardware implementations of the FFT [5][6][10] due to their regularity, simple control, pipelined operation, and low hardware resources usage; the R2$^2$SDF architecture is based on a radix-$2^2$ algorithm and it is suitable for FFT hardware [5][6][10][11] . Figure 3 shows the R2$^2$SDF architecture for a 64-point FFT. The R2$^2$SDF architecture uses two types of butterflies, which have a similar structure to the radix-2 butterfly, the R2$^2$SDF architecture has a resource usage which compares to the radix-4 algorithms [10]. From Figure 3 it can be seen that control is performed by a $\log_2(N)$-bit counter.

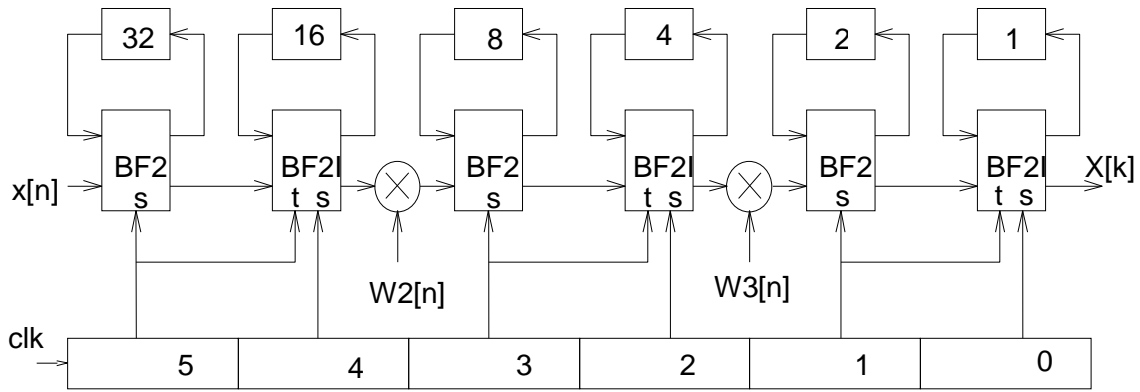

Figure 3: R2$^2$SDF architecture for 64-point FFT.

***OpenRISC processor and the Wishbone bus***

OpenRISC is a Reduced Instruction Set Computer (RISC) 32-bit soft-core processor designed by the community OpenCores, its architecture is described in a standard document [7]; also a synthesizable description in Verilog under the LGPL license is available through the OR1200 core [12]. OpenRISC allows the development of SOCs by using the interconnection bus Wishbone which is described in a standard document [8]. the OpenCores community has developed numerous cores with Wishbone connectivity such as Universal Asychronous Receiver Transmitter (UART), memory controller, Ethernet controller, timer controller, among others; however the OpenCores community has no DSP cores with Wishbone connectivity. Figure 4 shows a basic Wishbone interconnection between a master device and a slave device. Here, the master is either the OpenRISC processor or a bus controller; the slave is any Input/Output (I/O) device, coprocessor or hardware accelerator. According to the Wisbone specification [8] the signals in Figure 4 are described in Table 1. The DSP cores we designed are Wishbone compatible, and they use the basic connection depicted in Figure 4. The OpenCores community has developed some reference SoCs based on the OpenRISC processor which are FPGA-synthesizable; one of the simplest is MinSoC [13] which allows an easy and fast verification of the OpenRISC-based SoC with custom slave modules such as the DSP cores we designed.

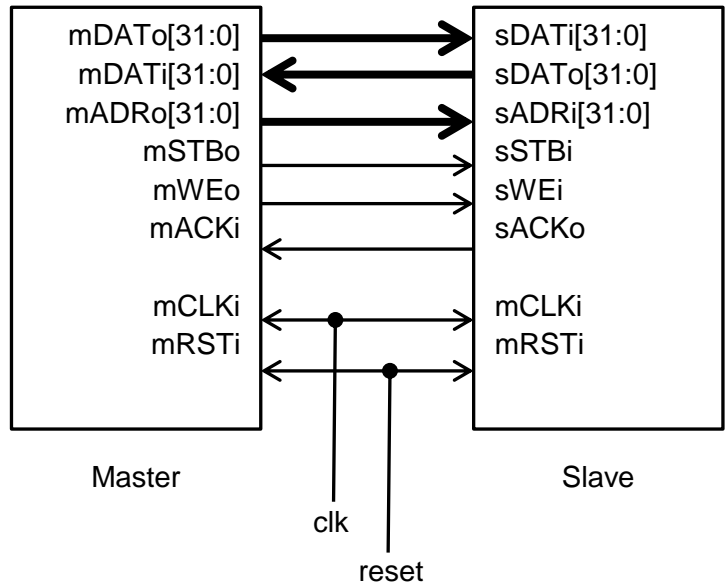

Figure 4: Wishbone bus basic connection.

| Signal | Description |
|---|---|
| mDATo/sDATi | The data bus from master to slave (write operations). |
| mDATi/sDATo | The data bus from slave to master (read operations). |
| mADRo/sADRi | The address bus from master to slave. |
| mSTBo/sSTBi | The chip select signal from master to slave. It is set by master during a write/read operation. |
| mWEo/sWEi | The write enable signal from master to slave. It is set by master during a write operation. |
| mACKi/sACKo | The acknowledgment signal from slave to master. It is set by slave after a successful write/read operation. |
| clk | The global clock signal. |
| reset | The global active high reset signal. |

Table 1: Signal description of Wishbone bus.

## DSP CORES ARCHITECTURE

In this section we describe the designed DSP cores and the slave interfaces with the Wishbone bus. For all DSP cores we used fixed point arithmetic, here the word width, bit growth, fractional part [2], filters order, and FFT length [1] are parameterizable features through VHDL generics. Each DSP core is composed of two functional unit, the processing unit and the slave interface unit; the processing unit performs the DSP operation according to the considered core and the slave interface unit is the Wishbone interface for the SoC connection.

**FIR filter core**

Figure 5 shows the block diagram of a FIR filter core parameterized with $N = 50$, word width of *M*=16 bits, and a bit growth of *G*=8 bits. In this case the processing unit was designed by using the transpose realization form shown in Figure 1. In Figure 5 the signal ports with no connection are Wishbone compatible signals for OpenRISC-based SoC integration. The remaining signal are described as follows: The input_signal/sdat_o pair is the port with the signal to be filtered and it has a width of *M+G* bits; the output_signal/sdat_i pair is the port with the filtered signal and it has a width of *M+G* bits; the enable/start pair is the port which enables the filtering process in the processing unit; the filter_coef/HQ pair is the port with the filter coefficients and it has a width of *NxM* bits; the Q/Q pair is the port with the number of fractional bits in the fixed-point representation of the filter coefficients and it has a width of 4 bits. Table 2 shows the register description of the Wishbone interface for the FIR filter core.

| Register | Address |
|---|---|
| FIR_CONTROL[0:0] | FIR_BASE + 0 |
| FIR_DATA[M+G-1:0] | FIR_BASE + 4 |
| FIR_STATUS[0:0] | FIR_BASE + 8 |
| FIR_Q[3:0] | FIR_BASE + 12 |
| FIR_COEFF | FIR_BASE + 16 |

Table 2: Register description of the FIR filter core.

FIR_DATA is a read/write register used to write/read the input/filtered sample. FIR_CONTROL is an only write register; when it is set by the user, the filtering process is started. FIR_STATUS is an only read register; it is set when the filtering process finishes. FIR_Q is an only write register; in this address the user writes the number of fractional bits of the fixed-point representation of the filter coefficients.

From the FIR_COEFF starts an addressing space composed of *N* consecutive 32-bit address positions where the user can write the 16-bit fixed-point filter coefficients starting from FIR_COEFF for $h[0]$ and finishing with FIR_COEFF+4x(N-1) for $h[N-1]$.

**IIR filter core**

Figure 6 shows the block diagram of a IIR filter core parameterized with $N_{sect} = 6$, word width of *M*=16 bits, a bit growth of *G*=8 bits, and *Q*=13 fractional bits. In this case the processing unit was designed by using a cascade of $N_{sect}$ pipelined SOS as the shown in Figure 2. In Figure 6 the signal ports with no connection are Wishbone compatible signals for OpenRISC-based SoC integration. The remaining signal are described as follows: The input_signal/sdat_o pair is the port with the signal to be filtered and it has a width of *M+G* bits; the output_signal/sdat_i pair is the port with the filtered signal and it has a width of *M+G* bits; the enable/start pair is the port which enables the filtering process in the processing unit; the enable_out/enable_in pair is the flag which signals the filtering process completion; the filter_coef/HQ pair is the port with the filter coefficients and it has a width of 6x*N*<sub>sect</sub>x*M* bits; the gain/gain pair is the port with the gain $\sqrt[N_{sect}]{G}$ for each SOS and it has a width of *M* bits; the en_out/en_out pair is the port with the number of used minus one sections from the $N_{sect}$ available sections and it has a width of 4 bits. In this case *Q* fractional bits are used for the fixed-point representation of the filter coefficients and the gain. Table 3 shows the register description of the Wishbone interface for the IIR filter core. IIR_DATA is a read/write register used to write/read the input/filtered sample. IIR_CONTROL is an only write register; when it is set by the user, the filtering process is started. IIR_STATUS is a write/read register; it is set when the filtering process finishes, it is cleared with a write operation. IIR_NSECT is an only write register; in this address the user writes the number of used minus one sections from the *N*<sub>sect</sub> available sections; IIR_GAIN is an only write register; in this address the user writes the gain $\sqrt[N_{sect}]{G}$ for each SOS by using fixed-point representation with *Q* fractional bits. From IIR_COEFF starts an addressing space composed of 6x*N*<sub>sect</sub> consecutive 32-bit address positions where the user can write the 16-bit fixed-point SOS coefficients starting from IIR_COEFF for $a_2$-1$^{st}$ SOS section, IIR_COEFF + 4 for $a_1$-1$^{st}$ SOS section, IIR_COEFF + 8 for $a_0$-1$^{st}$ SOS section, IIR_COEFF + 12 for $b_2$-1$^{st}$ SOS section, IIR_COEFF + 16 for $b_1$-1$^{st}$ SOS section, and IIR_COEFF + 20 for $b_0$-1$^{st}$ SOS section; and finishing with IIR_COEFF + 4x(6x$N_{sect}$-6) for $a_2$-$N_{sect}$$^{st}$ SOS section, IIR_COEFF + 4x(6x$N_{sect}$-5) for $a_1$-$N_{sect}$$^{st}$ SOS section, IIR_COEFF + 4x(6x$N_{sect}$-4) for $a_0$-$N_{sect}$$^{st}$ SOS section, IIR_COEFF + 4x(6x$N_{sect}$-3) for $b_2$-$N_{sect}$$^{st}$ SOS section, IIR_COEFF + 4x(6x$N_{sect}$-2) for $b_1$-$N_{sect}$$^{st}$ SOS section, and IIR_COEFF + 4x(6x$N_{sect}$-1) for $b_0$-$N_{sect}$$^{st}$ SOS section.

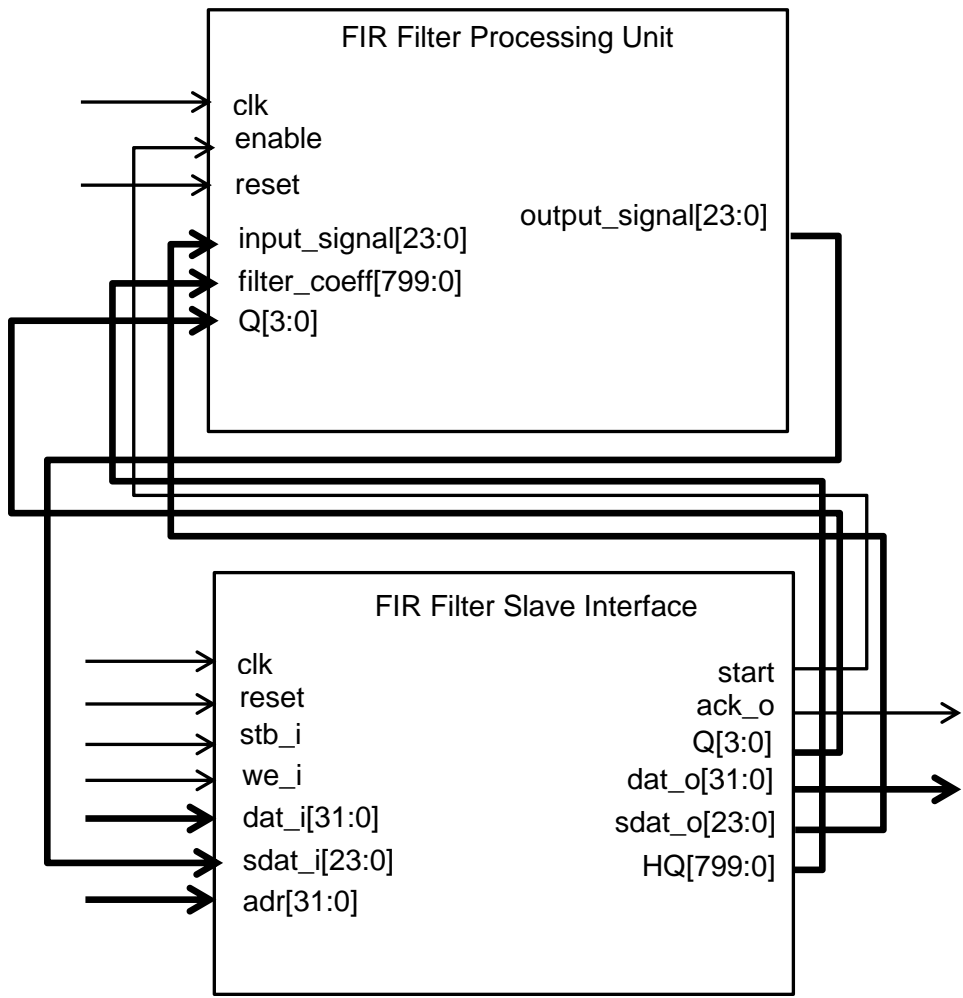

Figure 5: FIR Filter core.

| Register | Address |
|---|---|
| IIR_CONTROL[0:0] | IIR_BASE + 0 |
| IIR_DATA[M+G:0] | IIR_BASE + 4 |
| IIR_STATUS[0:0] | IIR_BASE + 8 |
| IIR_NSECT[3:0] | IIR_BASE + 12 |
| IIR_GAIN[15:0] | IIR_BASE + 16 |
| IIR_COEFF | IIR_BASE + 20 |

Table 3: Register description of the IIR filter core.

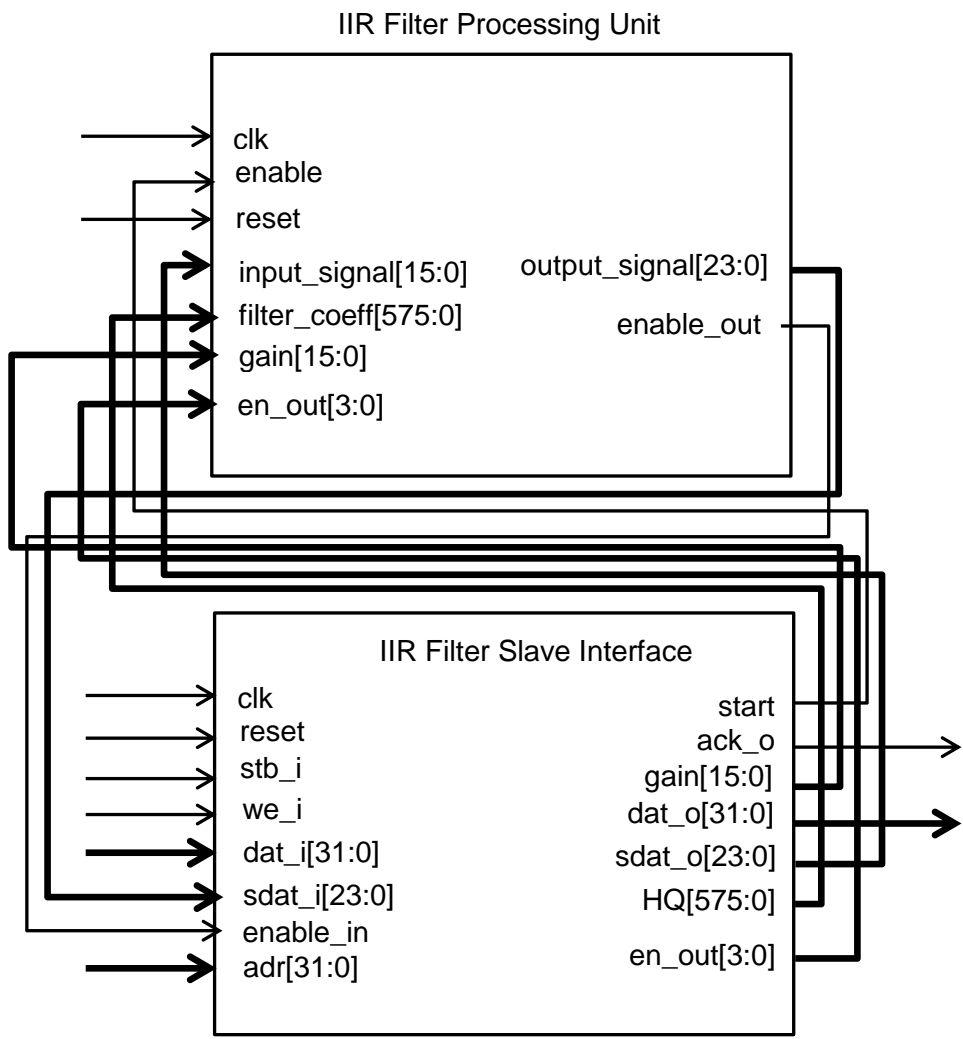

Figure 6: IIR Filter core.

**FFT core**

Figure 7 shows the block diagram of a FFT core parameterized with $N = 1024$, word width of *M=16* bits, and *Q=15* fractional bits. In this case the processing unit was designed by using the pipelined version of the R2$^2$SDF architecture developed in [11]. In Figure 7 the signal ports with no connection are Wishbone compatible signals for OpenRISC-based SoC integration. The remaining signal are described as follows: The Xinr/dat[15:0] pair is the port with the real part of the input sample and it has a width of *M* bits; Xini/dat[31:16] pair is the port with the imaginary part of the input sample and it has a width of *M* bits; the enable/fft_enable pair is the port which enables the FFT computing for each input sample in the processing unit; the Xoutr/sdat_i[15:0] pair is the port with the real part of the output sample and it has a width of *M* bits; Xouti/sdat_i[31:16] pair is the port with the imaginary part of the output sample and it has a width of *M* bits; the enable_out/fft_enable_in pair is the flag which signals the FFT process completion for each input sample; the frame_ready/fft_finish pair is the flag which signals the whole FFT process completion; the index/adr_fft pair is the bit-reversed address [1][2] in On-chip RAM where the processed sample is written. Table 4 shows the register description of the Wishbone interface for the FFT core. FFT_DATA is an only write register used to write the input samples. IIR_CONTROL is an only write register; when it is written by the user the processing unit and the status register are cleared; IIR_STATUS is an only read register; it is set when the whole FFT process finishes. From FFT_MEMORY starts an addressing space composed of *N* consecutive 32-bit address positions where the user can read the FFT results starting from FFT_MEMORY for $X[0]$ and finishing with FFT_MEMORY + 4x(*N-1)* for $X[N-1]$.

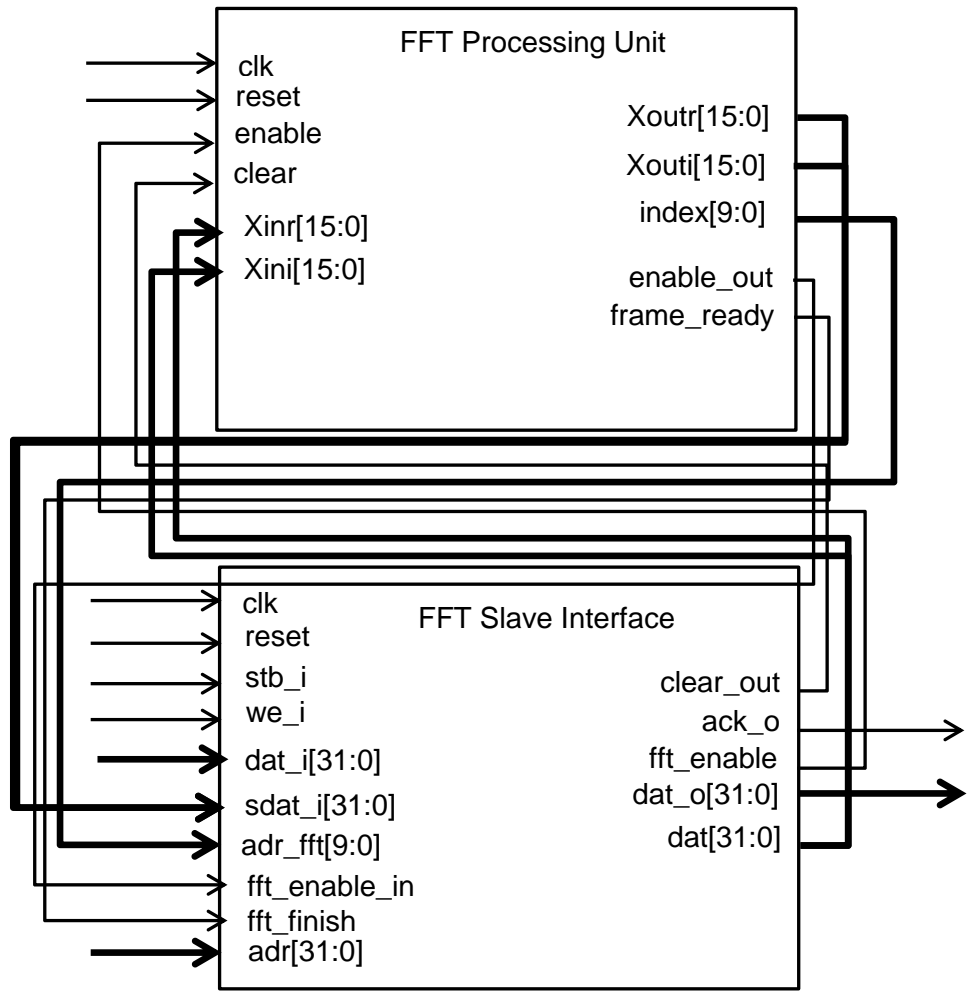

Figure 7: IIR Filter core.

| Register | Address |
|---|---|
| FFT_CONTROL[0:0] | FFT_BASE + 0 |
| FFT_DATA[31:0] | FFT_BASE + 4 |
| FFT_STATUS[0:0] | FFT_BASE + 8 |
| FFT_MEMORY | FFT_BASE + 12 |

Table 4: Register description of the FFT core.

## IN-SYSTEM HARDWARE VERIFICATION

The three DSP cores were integrated into an OpenRISC based SoC built from the reference design MinSoC [13]. The SoC was synthesized on the Altera FPGA device EP2S60F1020C4 included in the development board TREX-S2-TMB [14]. The accuracy of the cores was measured in terms of the Mean Squared Error (MSE) between frequency responses in DFT domain as shown in Eq. (4)

$$MSE = \sum_{k=0}^{N-1} |X[k] - \tilde{X}[k]|^2 \qquad (4)$$

Here $X[k]$ is the frequency response in the DFT domain when it is computed by simulation using double precision floating point arithmetic, and $\tilde{X}[k]$ is the frequency response of the core in DFT domain.

In the case of the FIR filter core we designed a 49-th order equiripple low-pass filter with cutoff frequencies 1.178097245096172 rad/s and 1.570796326794897 rad/s. The core was parameterized with a word width of 16 bit, a fractional part of 15 bits, and a bit growth of 8 bits. Figure 8 shows the magnitude frequency responses for the FIR filter core test.

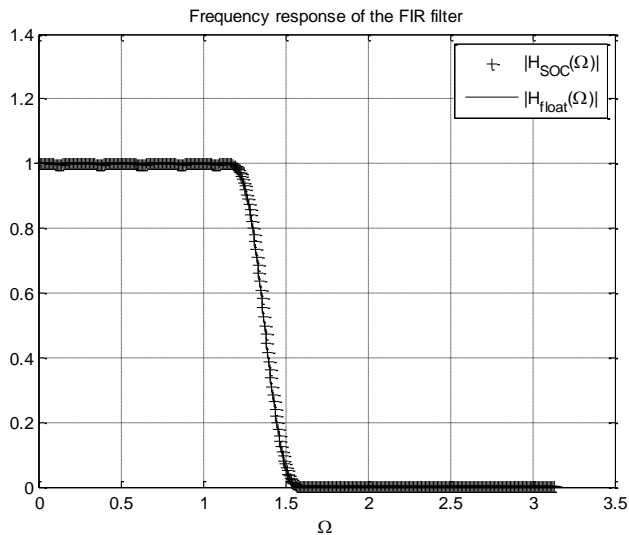

Figure 8: Frequency response of the FIR filter core.

In Figure 8 the continuous line depicts the magnitude frequency response of the FIR filter when it is computed by simulation using double precision floating point arithmetic. The frequency response of the FIR filter core was computed by getting the impulse response and taking its DFT, this is depicted with the dotted line. In this case the $MSE$ is $1.160378660241209 \times 10^{-4}$. Table 5 shows the synthesis report for the FIR filter core.

| Parameter | Value |
|---|---|
| Logic utilization | 27 % |
| Combinational ALUTs | 11,598 / 48,352 ( 24 % ) |
| Dedicated logic registers | 1,947 / 48,352 ( 4 % ) |
| Total block memory bits | 0 / 2,544,192 ( 0 % ) |
| DSP block 9-bit elements | 100 / 288 ( 35 % ) |
| Maximum operating frequency | 103.92 MHz |

Table 5: Synthesis report for the FIR core with $N = 50$.

The FIR filter core requires the 27% of the resources and reaches up a maximum operating frequency of 103.92 MHz.

In the case of the IIR filter core we designed a 12-th order Butterworth band-pass filter with cutoff frequencies 0.10625 rad/s, 0.11875 rad/s, 0.1025 rad/s, and 0.1225 rad/s. The core was parameterized with 6 SOS sections, a word width of 16 bit, a fractional part of 13 bits, and a bit growth of 8 bits. Figure 9 shows the magnitude frequency responses for the IIR filter core test.

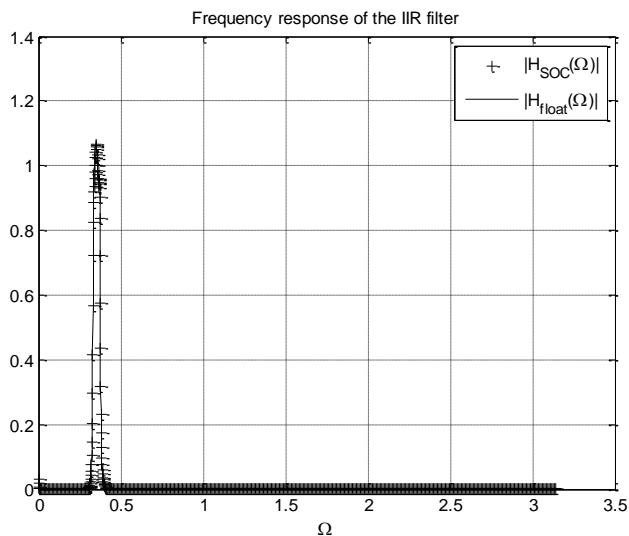

Figure 9: Frequency response of the IIR filter core.

In Figure 9 the continuous line depicts the magnitude frequency response of the IIR filter when it is computed by simulation using double precision floating point arithmetic. The frequency response of the IIR filter core was computed by getting the impulse response and taking its DFT, this is depicted with the dotted line. In this case the $MSE$ is $4.830547250436524 \times 10^{-5}$.

Table 6 shows the synthesis report for the IIR filter core.

| Parameter | Value |
| --- | --- |
| Logic utilization | 5 % |
| Combinational ALUTs | 2,201 / 48,352 ( 5 % ) |
| Dedicated logic registers | 504 / 48,352 ( 1 % ) |
| Total block memory bits | 0 / 2,544,192 ( 0 % ) |
| DSP block 9-bit elements | 288 / 288 ( 100 % ) |
| Maximum operating frequency | 85.81 MHz |

Table 6: Synthesis report for the IIR core with $N_{sect} = 6$.

The IIR filter core requires the 5% of the resources and reaches up a maximum operating frequency of 85.81 MHz.

The FFT core was parameterized with 1024 points, a word width of 16 bit, a fractional part of 15 bits, and a total gain of $2^{-4}$. Figure 10 shows the frequency responses for the FFT core test.

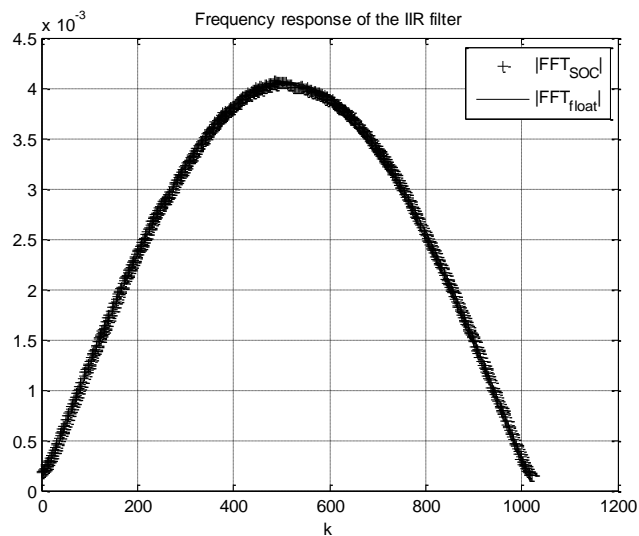

Figure 10: Frequency response of the FFT core.

In this case we computed the FFT for the signal $x[n] = \{-0.002098083496094,\ 0.001953125, 0, 0, \ldots, 0\} \vee 0 \leq n < 1024$. In Figure 10 the continuous line depicts the FFT computed by simulation using double precision floating point arithmetic; the dootted line depicts the FFT computed by the core. In this case the $MSE$ is $1.643710888516962 \times 10^{-9}$.

Table 7 shows the synthesis report for the FFT core.

| Parameter | Value |
|---|---|
| Logic utilization | 69 % |
| Combinational ALUTs | 1,085 / 48,352 ( 2 % ) |
| Dedicated logic registers | 33,095 / 48,352 ( 68 % ) |
| Total block memory bits | 73,728 / 2,544,192 ( 3 % ) |
| DSP block 9-bit elements | 32 / 288 ( 11 % ) |
| Maximum operating frequency | 115.3 MHz |

Table 7: Synthesis report for the FFT core with $N = 1024$.

Tthe FFT core requires the 69% of the resources and reaches up a maximum operating frequency of 115.3 MHz.

Table 8 shows the synthesis report for the whole OpenRISC-based DSP SoC.

| Parameter | Value |
|---|---|
| Logic utilization | 100 % |
| Combinational ALUTs | 33,568 / 48,352 ( 69 % ) |
| Dedicated logic registers | 39,428 / 48,352 ( 82 % ) |
| Total block memory bits | 494,464 / 2,544,192 ( 19 % ) |
| DSP block 9-bit elements | 288 / 288 ( 100 % ) |
| Maximum operating frequency | 55.05 MHz |

Table 8: Synthesis report for the OpenRISC-MinSoC-based DSP SoC.

In this case the OpenRISC-based DSP SoC requires the 100% of the resources and reaches up a maximum operating frequency of 55.05 MHz. The constraint in operating frequency is due to the MinSoC SoC and not the DSP Cores.

## CONCLUSION

We designed the DSP Cores FIR filter, IIR filter, FFT which are compatible with the Wishbone bus allowing the construction of DSP SoC based on the OpenRISC processor. The three DSP cores are parameterizable through VHDL generics and they have easy to use hardware/software interfaces. The three DSP cores we designed are the only of their kind in the OpenCores community because of the broad DSP functions availability, the Wishbone compatibility, the flexibility, and speed performance. The three cores have been tested on Altera FPGA devices such Cyclone II and Stratix II.

## ACKNOWLEDGEMENTS

Juan Camilo Valderrama-Cuervo thanks Prof. López-Parrado for the given support and teachings. Alexander López-Parrado thanks Colciencias for the scholarship, and he also thanks Universidad del Quindío for the study commission.